# Dynamic and Static Transmission Electron Microscopy Studies on Structural Evaluation of Au nano islands on Si (100) Surface

A. Rath[1], R. R. Juluri[1] and P. V. Satyam[1,*]

[1] Institute of Physics, Sachivalaya Marg, Bhubaneswar - 751005, India

**Abstract:**

Transmission electron microscopy (TEM) study on morphological changes in gold nanostructures deposited on Si (100) upon annealing under different vacuum conditions has been reported. Au thin films of thickness ~2.0 nm were deposited under high vacuum condition (with the native oxide at the interface of Au and Si) using thermal evaporation. In-situ, high temperature (from room temperature (RT) to 850ºC) real time TEM measurements showed the evaluation of gold nanoparticles into rectangular/square shaped gold silicide structures. This has been attributed to selective thermal decomposition of native oxide layer. Ex-situ annealing in low vacuum ($10^{-2}$ mbar) at 850ºC showed no growth of nano-gold silicide structures. Under low vacuum annealing conditions, the creation of oxide could be dominating compared to the decomposition of oxide layers resulting in the formation of barrier layer between Au and Si.

# 1. Introduction

The major issue in nanotechnology is the development of conceptually simple synthesis technique for the mass fabrication of nano scale structures. At this level, the conventional top-down approach becomes expensive and complicated process. One of the alternative methods is to make use of the self assembly growth procedure. Understanding of self assemble growth of nanostructures require a detailed knowledge that is based on the principle of microscopic pathways of diffusion, nucleation and aggregation. One of the challenges in the self assembly growth process is to understand the nucleation process with atomic scale spatial resolution in real time sequence. Dynamic *in-situ* TEM studies involving a capability of variable temperature and with tens of millisecond time resolution play a key role in understanding the growth kinetics. In this work, we present on the formation of nano gold silicides at an early stage in case of a 2.0 nm thick gold film deposited on Si (100) with native oxide at the interface. Our results show that the oxide at the interface inhibits the gold reaction with silicon substrate and this resulting in spherical gold nanostructures for annealing in low vacuum conditions.

It is well known that metal nano-particles exhibit unique electronic, magnetic, photonic and catalytic properties resulting in the preparation of new materials for energy storage, photonics, communications and sensing applications [1, 2]. For tailoring the physical and chemical properties and their uses, the shape, size and composition play an important factor [3]. However, controlling the shape of nano particles grown by thin film technologies is generally difficult due to its dependence on kinetic and thermodynamic parameters which are stochastic in nature [4]. This approach usually requires thermal activation through heating the support. In this work, we deal with the real time observation of nano – gold silicide structure formation at high temperature and various vacuum conditions. Gold nano crystals with various shapes (rods, spheres and squares etc) have been reported using various methods [5 – 7]. In the present paper, we report the role of vacuum conditions on the evaluation of various morphological changes that occur for the nano gold particles on Si (100) annealed at high temperatures. Our results also show the enhancement in the thermal decomposition of underlying native/thermal oxide layer on Si(100) due to the presence of gold nanostructures grown on top of the native oxide. Some important aspects of such decomposition on Si (111) and Si(100) systems have been reported [9]. At elevated temperature and high vacuum conditions, silicon oxide thin films known to decompose by the overall reaction Si (s) + $SiO_2$ (s) —> 2 SiO (g) [10, 11]. In this process, voids are formed in the oxide layer exposing the substrate silicon. It has been reported that presence of a metallic thin film enhances such decomposition [12-14]. When such decomposition takes place during low vacuum conditions, one has to deal with simultaneous growth of oxide layers as well. This would contribute to play important role in controlling the inter-diffusion and reaction of gold with silicon. These effects are also reported in this work for a 2.0 nm gold deposited on native oxide silicon substrate.

## 2. Experimental Details

Gold films of ~2.0 nm thickness were deposited by thermal evaporation method under high vacuum ($\approx 4\times10^{-6}$ mbar) conditions on n-type Si (100) substrates. For these systems, thin native oxide with thickness of ~2.0 nm has been observed using cross-sectional TEM measurements. Planar TEM specimens were prepared from these samples. In-situ heating experiments were carried out by using a GATAN hot-stage (Model 628 UHR single tilt heating holder) TEM holder. Real time measurements were acquired using a CCD camera (GATAN 832) in which real time movies can also be recorded with 25 frames $s^{-1}$ (40 ms resolution). The sample was annealed inside the TEM chamber (HV) up to 850ºC (system-A). Another *as-*

*deposited* sample was annealed in a low vacuum furnace ($\approx 10^{-2}$ mbar) at $\approx 850°$C for 30 minutes (system-B). Planar TEM samples were prepared out of that annealed specimen for further TEM measurements with 200 keV electrons (2010, JEOL HRTEM).

## 3. Results and discussion

Fig. 1(a) depicts a bright field (BF) planar TEM micrograph for ~ 2.0 nm thick Au film grown on Si (100) substrate with a 2.0 nm thin native oxide at the interface between the Au nanostructures and the substrate. Irregular shaped gold nanostructures were seen with 18% area coverage of gold nanostructures. The selected area diffraction pattern (SAD) taken on a group of nanostructures confirms the polycrystalline nature of the gold films and along with the reflections of substrate silicon (fig. 1(b)). The above system was annealed from (room temperature) RT to 850ºC inside TEM column using the single tilt heating stage (system-A). We have observed that at 600ºC, the surface morphology was almost similar to that of *as deposited* system (as shown in Fig 1(a)). We noted the start of agglomeration of later diffusion of gold particles (kind of Oswald ripening) at 700ºC happening at random sites and this process increased (i.e. the number of clusters) with increasing temperature. At higher temperatures, i.e., around 850ºC, interestingly, desorption of the presumably gold silicide structures leading to the formation of hole like structures inside big clusters of gold silicide structures has been noted. This can be attributed to the melting induced desorption of gold from these silicide island structures [7]. Interesting observation of nanosized gold particle movement and leading to the formation of nanosized ordered (square or rectangular) gold silicide structures has been seen. The contrast confirms the formation of silicide structures and later on, this is confirmed by the in-situ SAD pattern. This was explained based on the selective decomposition of native oxide at high vacuum and temperatures and there upon the gold diffuses towards the exposed silicon surface to form rectangular gold silicide structures [7]. In between these rectangular structures, un-reacted gold particles were still present (dark contrast) (Fig. 2). The variation of contrast compared to that of un-reacted gold, indicates the formation of gold silicide. Fig. 1(d) depicts selected area diffraction (SAD) pattern where the ring (arrow marked) corresponds to the d-spacing of 0.253 nm. It matches with both the $Au_5Si_2$ and $Au_3Si$ phase of gold silicide [15]. It is well known that, at high temperatures and good enough vacuum/oxygen free conditions, oxide layer undergoes thermal decomposition. Presence of gold acts as a catalyst for the decomposition to take place. According to Dallaporta et al [13], heat of reaction between gold and $SiO_2$ being positive [16], lack of reactivity requires that Au has to reach the $Si/SiO_2$ interface so that it forms gold silicide and enhances oxide decomposition process.

Fig. 2 shows the real time BF TEM image depicting the growth of nano rectangular/square shaped structures at the expense of gold nano particles. The Images are taken after the temperature was stabilized at 850°C. We presume the time at which the image shown in fig. 2(a) as starting time (t=0.0s). At t=8.0s (fig. 2(b)) there was no change in the morphology of the two nano particles (inside white circle). At t= 8.64s, they joined together (fig. 2(c)). As time progressed, it tried to rearrange itself to get minimum energy configuration [17, 18] which resulted in formation of rectangular/square shaped silicide structures (fig. 2(d) and 2(e)). Interestingly, we do not observe any further agglomeration after t= 12.12s (fig. 2(f)), (even after waiting for more than 20 min). The reason to such phenomena is not known yet as per our knowledge.

In fig. 3, BF TEM image depicting the real time morphological changes of gold particle surrounded by group of particles during in situ heating (stabilized at the temp of 850°C). Here also, we presume the time at which the image shown in fig. 3(a) as starting time (t=0.0s). At t=0.0s, particles denoted by legends A, B, C, D, E, F, G and H were present around the region of interest (ROI) structure (inside dotted white line: Fig. 3)) at a distance of 18.4 nm, 67.6nm, 56.3 nm, 28.7 nm, 55.4 nm, 107.4 nm, 132.9 nm and 32.2 nm respectively (fig. 3(a)). After 6.32s, particle 'A' diffused into the ROI structure which resulted in increase of the size of the ROI structure (fig. 3(b)). All the above 8 particles diffused into the ROI structure in 19.08s (fig. 3(c)-3(h)). The contrast of the growing ROI structure and the gold particle confirms that gold from the particles diffuses into the Au–Si alloy (ROI) in a process more like Oswald ripening, similar to the results discussed by Kim *et al* [19]. Particles from all direction of the ROI structure are diffusing due to the square symmetry of the Si (100). Umananda et al showed the uni-directional growth of gold silicide rods which is attributed to the selective decomposition of oxide growth on the Si(110) surface [5]. After each intake of the particle, there was a increase in size of the ROI structure. After that, particles are not further diffusing and one can see the formation of hole like structures in the ROI structure due to the desorption of gold (fig. 3(i)). Interestingly, no growth of gold silicide structure has been observed after the start of formation of hole. In each image frame, the already nucleated island with holes is present near the ROI structure without interacting with the nearest particles. At t= 49.92s, more holes were formed in the ROI structure. Even after waiting for more than 30 min, there was no change in morphology of the structure (fig. 3(i)). It shows that the particles become stable after some critical size. All the images shown in figure 2 and figure 3 are reproduced from the original video.

We now discuss the role of vacuum condition on decomposition and void formation process. In a bid to study this aspect, the *as-deposited* sample was ex-situ annealed at 850°C for half an hour inside the low vacuum furnace (system-B). Vacuum was kept at about $10^{-2}$ mbar (low vacuum obtained using a rotary pump). The system was allowed to cool down to room temperature (RT) for further TEM measurements. Fig. 4(a) shows the bright field transmission electron micrographs taken at RT after vacuum annealing of 2nmAu/SiO$_2$/Si(100) system at

850°C. Interestingly, spherical particles of Gold are observed with a monotonous size distribution. It is very interesting that even after annealing at such high temperatures, formation of aligned structures were not observed as in Fig. 1(c). As explained earlier, at such high temperatures, under high vacuum conditions (i.e., in the absence of $O_2$), thin oxide layers undergo a reaction $Si + SiO_2 \longrightarrow 2SiO$, where SiO is a volatile product at high temperatures. This results in selective removal of oxide layer, leading to the growth of aligned structures. But, when the vacuum level is lowered to about $10^{-2}$ mbar (and the vapor pressure of SiO being close to $10^{-1}$ mbar at $1200^{o}C$ [20]), rate of evaporation of volatile SiO will be affected. Also, at lower vacuum, mean free path of atoms/molecules present in the chamber reduces to fraction of a centimeter (which is of the order of kilo meters under UHV conditions) and the time required for the formation of a monolayer decreases to fraction of a milli second (from an hour for UHV). Thus, formation of a monolayer of external impurities (redeposition of $SiO_2$ here) takes much lesser time [21]. Thus, if the rate of redeposition of oxide exceeds the rate of decomposition, proper selective decomposition of oxide layer might not be possible. As a result, it hinders the growth of any gold reaching the surface silicon in forming aligned silicide structures. The SAD did not show any reflection of gold silicide phase except the signals of polycrystalline gold and the substrate silicon (fig. 4(b)).

## 4. Conclusions

2.0 nm Au were deposited on a Si(100) using thermal evaporation method (with native oxide) Planar TEM samples were prepared and loaded in a hot-stage holder for in-situ TEM measurements. Aligned nanostructures were observed at elevated temperature. Whereas for similar system, upon annealing externally in low vacuum condition leads to the formation of spherical nano gold particle (no formation of aligned gold nano-structures) establishing the fact that vacuum level plays an important role in the selective decomposition process.

**Figure Captions**

**Fig 1:** (a) 2nm As deposited thermally grown on Si(100) showing nanostructures (b) Corresponding SAD pattern showing the reflection of Au and silicon (c) Bright Field transmission electron micrograph at 850ºC (system-A) (d) Corresponding selected area diffraction pattern and the ring (arrow marked) indicates the alloy formation.

**Fig 2:** Real time bright field tem image depicting the growth of a nano rectangle at the expense of gold nano particles (at 850ºC)

**Fig 3:** Bright field TEM image depicting the real time morphological changes during in situ heating (stabilized at the temp of 850ºC)

**Fig 4:** (a) 2nm Au/SiO$_2$/Si(100) was ex-situ annealed (at 850º C) under low vacuum (system-B) and then seen in TEM at RT (b) corresponding SAD showing the reflection of gold and silicon

**Fig 1: Rath et al**

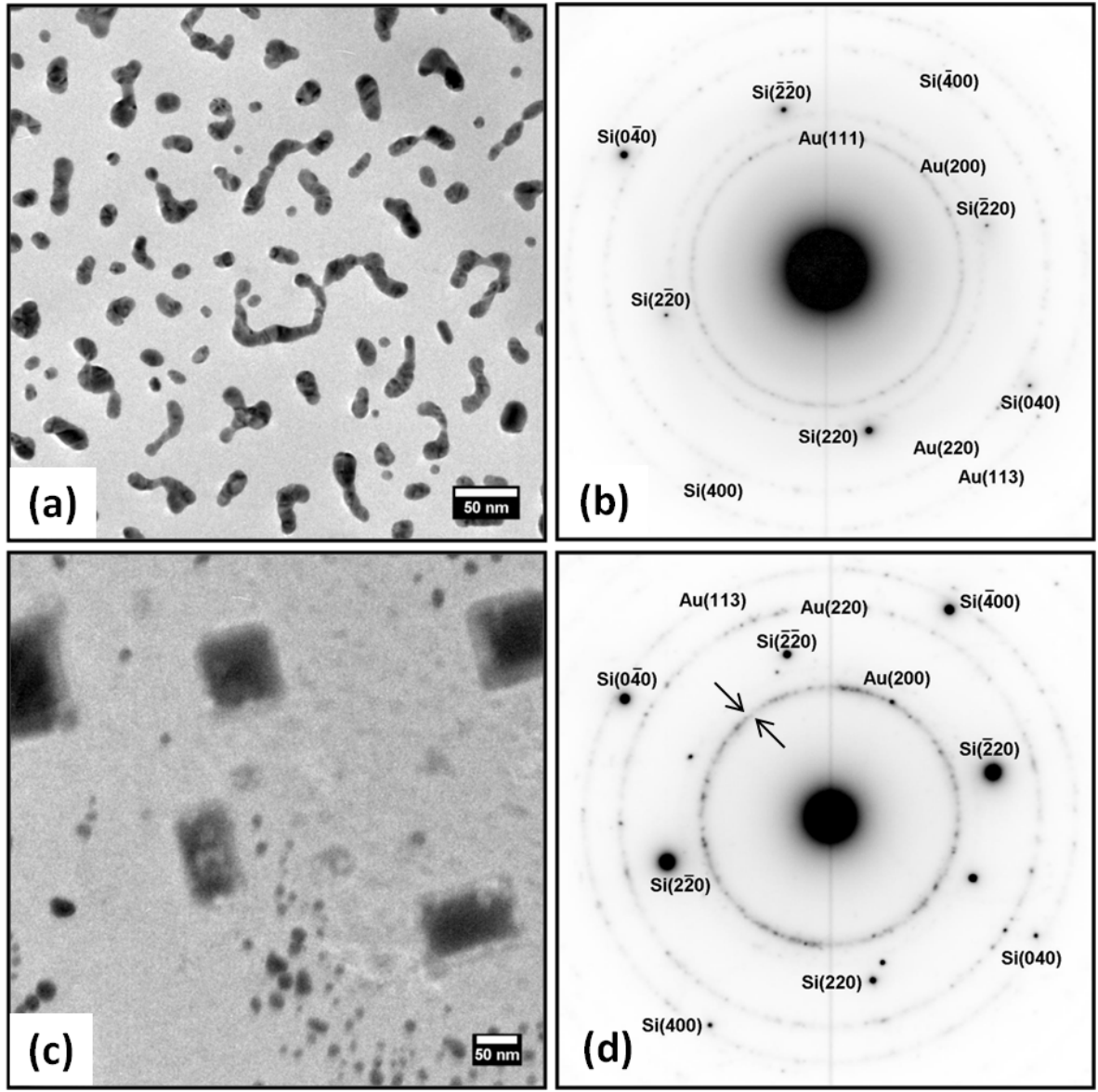

**Fig 2: Rath et al**

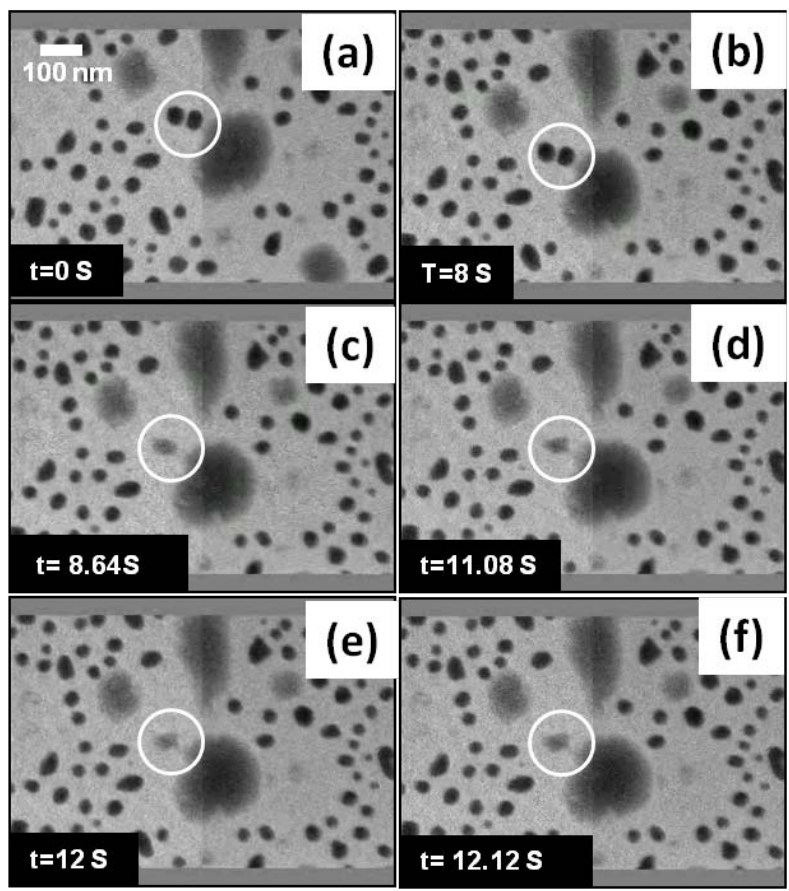

**Fig 3: Rath et al**

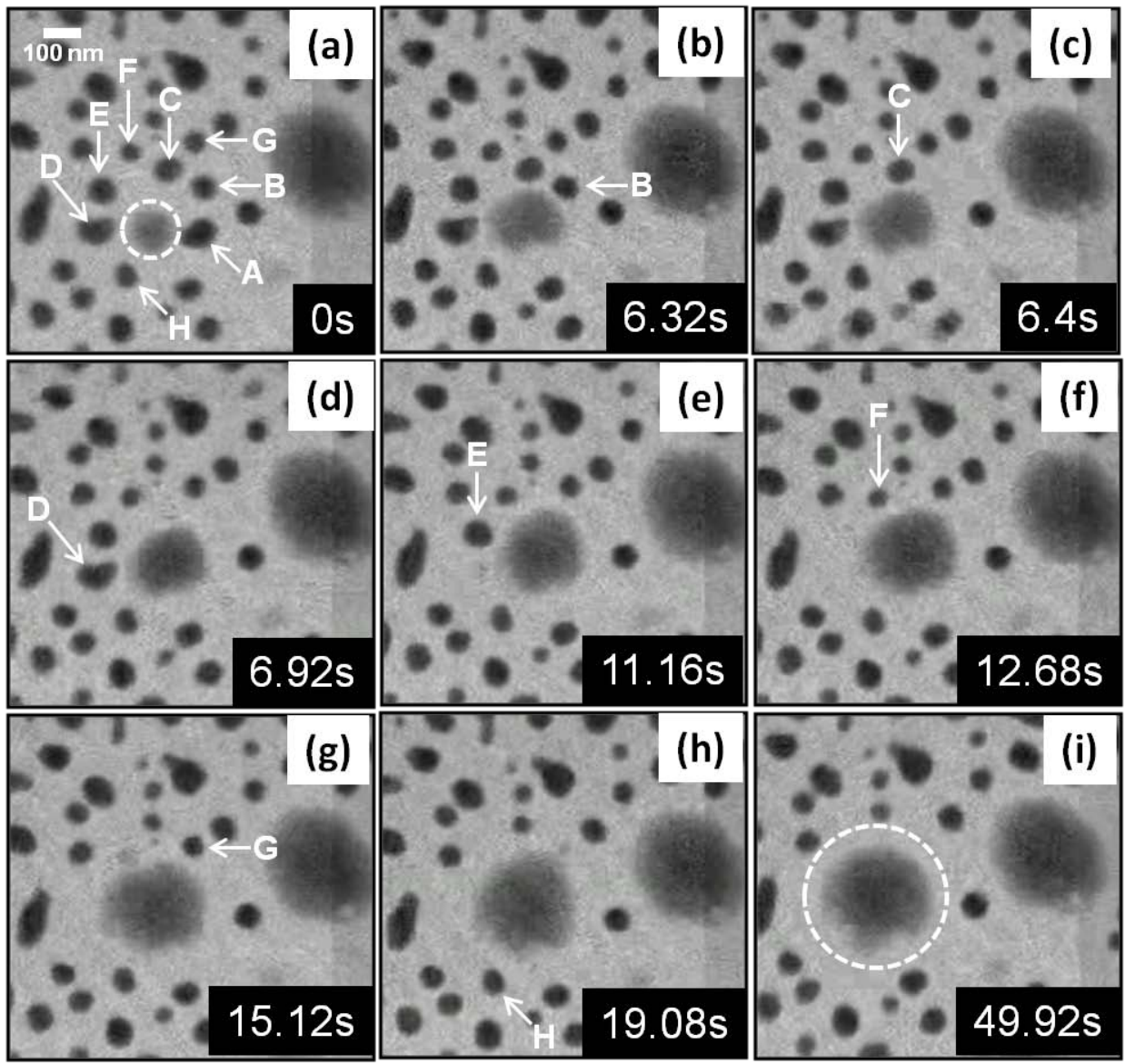

**Fig 4: Rath et al**

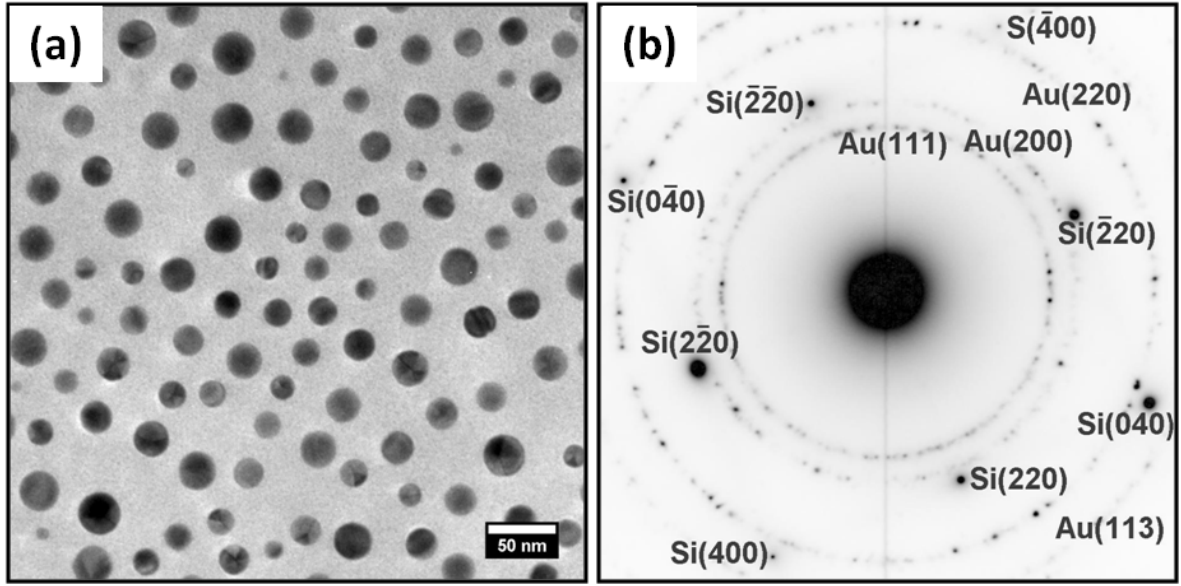